\documentclass[twocolumn, twocolappendix]{aastex631}

\usepackage{amsmath, mathtools, amsthm, amssymb}
\usepackage{enumitem}
\usepackage[toc,page]{appendix}
\hypersetup{colorlinks=true, allcolors=cyan}
\usepackage{soul}
\usepackage{graphicx}
\usepackage{appendix}
\graphicspath{ {./} }
\usepackage{xcolor}

\newcommand{\mstellar}{\ensuremath{M_{*}}}
\newcommand{\mhalo}{\ensuremath{M_{\rm halo}}}
\newcommand{\msol}{\ensuremath{\rm{M_{\odot}}}}
\newcommand{\mratio}{\ensuremath{\log(M_{\rm halo}/M_{*})}}
\newcommand{\elower}{\ensuremath{e}}

\usepackage[super]{nth}

\received{XX}
\revised{YY}
\accepted{ZZ}
\submitjournal{ApJ}

\begin{document}

\title{Star formation history and transition epoch of  cluster galaxies based on the Horizon-AGN simulation}


\author{Seyoung Jeon}
\affil{Department of Astronomy and Yonsei University Observatory, 50 Yonsei-ro, Seodaemun-gu, 
Seoul 03722, Republic of Korea}

\author{Sukyoung K. Yi}
\affil{Department of Astronomy and Yonsei University Observatory, 50 Yonsei-ro, Seodaemun-gu, 
Seoul 03722, Republic of Korea}

\author{Yohan Dubois}
\affiliation{Institut d’ Astrophysique de Paris, Sorbonne Universités, et CNRS, UMP 7095, 98 bis bd Arago, 75014 Paris, France}

\author{Aeree Chung}
\affil{Department of Astronomy and Yonsei University Observatory, 50 Yonsei-ro, Seodaemun-gu, 
Seoul 03722, Republic of Korea}

\author{Julien Devriendt}
\affiliation{Department of Physics, University of Oxford, Keble Road, Oxford OX1 3RH, UK}

\author{San Han}
\affil{Department of Astronomy and Yonsei University Observatory, 50 Yonsei-ro, Seodaemun-gu, 
Seoul 03722, Republic of Korea}

\author{Ryan A. Jackson}
\affil{Department of Astronomy and Yonsei University Observatory, 50 Yonsei-ro, Seodaemun-gu, 
Seoul 03722, Republic of Korea}

\author{Taysun Kimm}
\affil{Department of Astronomy and Yonsei University Observatory, 50 Yonsei-ro, Seodaemun-gu, 
Seoul 03722, Republic of Korea}

\author{Christophe Pichon}
\affiliation{Institut d’ Astrophysique de Paris, Sorbonne Universités, et CNRS, UMP 7095, 98 bis bd Arago, 75014 Paris, France}
\affiliation{IPhT, DRF-INP, UMR 3680, CEA, L’Orme des Merisiers, Bât 774, 91191 Gif-sur-Yvette, France}

\author{Jinsu Rhee}
\affil{Department of Astronomy and Yonsei University Observatory, 50 Yonsei-ro, Seodaemun-gu, 
Seoul 03722, Republic of Korea}


\begin{abstract}
Cluster galaxies exhibit substantially lower star formation rates than field galaxies today, but it is conceivable that clusters were sites of more active star formation in the early universe.
Herein, we present an interpretation of the star formation history (SFH) of group/cluster galaxies based on the large-scale cosmological hydrodynamic simulation, Horizon-AGN.
We find that massive galaxies in general have small values of e-folding timescales of star formation decay (i.e., ``mass quenching'') regardless of their environment, whilst low-mass galaxies exhibit prominent environmental dependence.
In massive host halos (i.e., clusters), the e-folding timescales of low-mass galaxies are further decreased if they reside in such halos for a longer period of time.
This ``environmental quenching'' trend is consistent with the theoretical expectation from ram pressure stripping.
Furthermore, we define a ``transition epoch'' as where cluster galaxies become less star-forming than field galaxies.
The transition epoch of group/cluster galaxies varies according to their stellar and host cluster halo masses.
Low-mass galaxies in massive clusters show the earliest transition epoch of $\sim 7.6$ Gyr ago in lookback time. However, it decreases to $\sim 5.2$ Gyr for massive galaxies in low-mass clusters.
Based on our findings, we can describe cluster galaxy's SFH with regard to the cluster halo-to-stellar mass ratio.
\end{abstract}

%

\section[]{Introduction}
\label{sec:introduction}

During recent decades, an enormous amount of information on galaxies has been gathered.
Despite the diversity and complexity of observed galaxy properties, many interesting statistics and relationships have emerged through numerous observational surveys.
Galaxies are often separated into red and blue galaxies on the color-magnitude plane, forming the red sequence and the blue cloud \citep[e.g.,][]{Strateva2001, Baldry2004}.
For example, the colors and the morphology of galaxies have a robust correlation with the environment \citep[e.g.,][]{Dressler1980, Hogg2003}.
While the majority of galaxies in the Universe are ``live'' and star-forming, most member galaxies of massive clusters are mysteriously ``red and dead'' with significantly depressed star formation rates (SFR) \citep[e.g.,][]{Balogh1999, Boselli2006}.

Various quenching processes have been proposed to explain the aforementioned diversities (in this study, ``quenched" refers to the cessation as well as the overall decline of star formation in the long-term timescale) and these processes are classified as internal and external.
Internal quenching regulates star formation mainly via feedback processes inside the galaxy.
For example, strong energy generated by an active galactic nucleus (AGN) heats the surrounding cold gas and directly transfers momentum, inhibiting the inflow of cold gas and regulating star formation, especially in the case of massive galaxies \citep[][]{Croton2006, Rafferty2006, Fabian2012, Peirani2017}.
In addition, stellar feedback can suppress star formation in various ways and is suggested to be the main quenching mechanism in low-mass galaxies in which gas is bound in a shallow potential well.
Supernova explosions provide energy and momentum to the surrounding interstellar medium (ISM), which can disturb cold gas and prevent it from collapsing \citep[][]{Larson1974, Dekel1986, Governato2010}.
Moreover, ultraviolet (UV) radiation generated by young massive stars contributes to star formation quenching through photoionization, which ionizes the surrounding cold gas and increases the gas pressure, or by direct radiative pressure \citep[][]{Whitworth1979, Larson1971, Hopkins2020}.

Alternatively, environmental effects could affect the SFR of galaxies in dense regions, where massive dark matter halos are formed.
For the galaxies orbiting in the deep gravitational potential well of a massive halo, their baryon and dark matter components are disrupted by tidal forces \citep[i.e., tidal stripping;][]{Richstone1976, Gnedin2003, Smith2016}.
When a galaxy moves in a cluster environment, the hot intracluster medium (ICM) can strip the ISM of the galaxy which is the fuel for star formation \citep[i.e., ram pressure stripping;][]{Gunn1972, Quilis2000, Chung2007, Tonnesen2007}.
Otherwise, the ICM can suppress hot gas cooling which consequently reduces cold gas supply to the galaxy \citep[i.e., starvation or strangulation;][]{Larson1980, Balogh2000}.
In dense regions, galaxies experience fly-by encounters and even mergers with nearby galaxies, resulting in the distortion of galactic structures \citep[i.e., harassment and mergers;][]{Moore1996, Moore1998, Yi2013, Smith2015, Sheen2016}.
Even in low-mass host halos, galaxies may already experience environmental effects before infalling into a cluster halo \citep[i.e., preprocessing;][]{Mihos2004, DeLucia2012, Han2018, Jung2018}.

Various quenching processes act with different timescales and magnitudes and are accumulated throughout the history of a galaxy's evolution \citep[][]{Iyer2020}.
Therefore, the star formation history (SFH) contains cumulative information about the quenching processes a galaxy has experienced.
As galaxies with different internal properties and external environments may have taken distinct evolutionary paths, we can connect their physical properties and SFHs.
Massive and quiescent galaxies are believed to have experienced an early vigorous star formation phase and subsequent rapid quenching, whereas low-mass galaxies are thought to have a relatively extended SFH \citep[e.g.,][]{Thomas2005, Ellison2018}.
Likewise, cluster galaxies can have different SFHs compared with field galaxies.
Although cluster galaxies are redder and more quenched than field galaxies at the present epoch, increased fractions of blue galaxies in high-z clusters exist \citep[][]{Butcher1978, Alberts2013}.
For example, it has been found that at $z\sim1$ cluster galaxies have comparable specific SFR (sSFR) and quenched fractions to field galaxies \citep[][]{Lee2015, Jian2020}.
Moreover, the local SFR-density relation seems to reverse at high redshifts.
This implies that there was an epoch in which there was no difference in the SFR between the field and cluster galaxies \citep[][]{Elbaz2007, Cooper2008, Hwang2019}.
This ``transition epoch'', when derived from the instantaneous SFR or quenching efficiency either through observations or simulations, seems to converge at $z\sim1$ \citep[][]{Alberts2013, Brodwin2013, Pintos-Castro2019, Hwang2019, Williams2022}.
On the other hand, considering that there have been many studies indicating various quenching timescales and processes depending on a galaxy's internal properties and environments, it would be mysterious if the transition epoch is universal.
Therefore, it is crucial for our understanding of galaxy formation and evolution to pin down such epochs and identify their driving mechanisms.

Numerical simulations have emerged as effective tools for studying the history of galaxy evolution.
Since the advent of large-scale N-body simulations \citep[][]{Springel2005}, cosmological simulations have become dramatically more sophisticated: to name only a few, Horizon-AGN \citep[][]{Dubois2014}, Illustris \citep[][]{Vogelsberger2014b}, EAGLE \citep[][]{Schaye2015, Crain2015}, Mufasa \citep[][]{Dave2017}, IllustrisTNG \citep[][]{Pillepich2018}, SIMBA \citep[][]{Dave2019}, and New-Horizon \citep[][]{Dubois2021}.
Along with the advances in our understanding of baryonic physics, these simulations have succeeded in reproducing some key properties of galaxies.

We use Horizon-AGN to investigate star formation quenching histories and assess the level of the reproduction of observed galaxy properties using the simulation.
We analyze the quenched states and SFHs of simulated galaxies in various host halo environments and attempt to assess the significance of the internal and external quenching processes on the SFHs of galaxies.
In addition, we determine whether there is an epoch where the star formation rates of cluster and field galaxies were equal.

%

\section[]{Methodology}
\label{sec:Method}

\subsection{Numerical Simulation}
\label{sec:Method-Num}

We use the large-scale cosmological hydrodynamic simulation, Horizon-AGN \citep[][]{Dubois2014}, which was performed using the adaptive mesh refinement code, RAMSES \citep[][]{Teyssier2002}.
Horizon-AGN uses WMAP7 values to set the cosmological parameters and includes subgrid astrophysics such as gas cooling due to primordial and metallic species and heating by the background UV radiation, star formation, stellar and AGN feedback \citep[][]{Dubois2014}.
The simulation volume is a $100\,{h^{-1}}\,{\rm Mpc}$ on a side cube in comoving scale and contains $1024^3$ dark matter particles.
The best spatial resolution is $\approx 1\,{\rm kpc\,(physical)}$ and the mass resolutions of the dark matter and star particles are $8\times 10^7\ \msol$ and $2 \times 10^6\ \msol$, respectively.
For a more detailed description of Horizon-AGN, refer to \cite{Dubois2014}. It has been demonstrated by \cite{Kaviraj2017} that Horizon-AGN shows reasonable agreement with key observational data such as mass function and cosmic star formation history, which are crucial in this study.

\begin{figure}[h]
\centering
\includegraphics[width=0.45\textwidth]{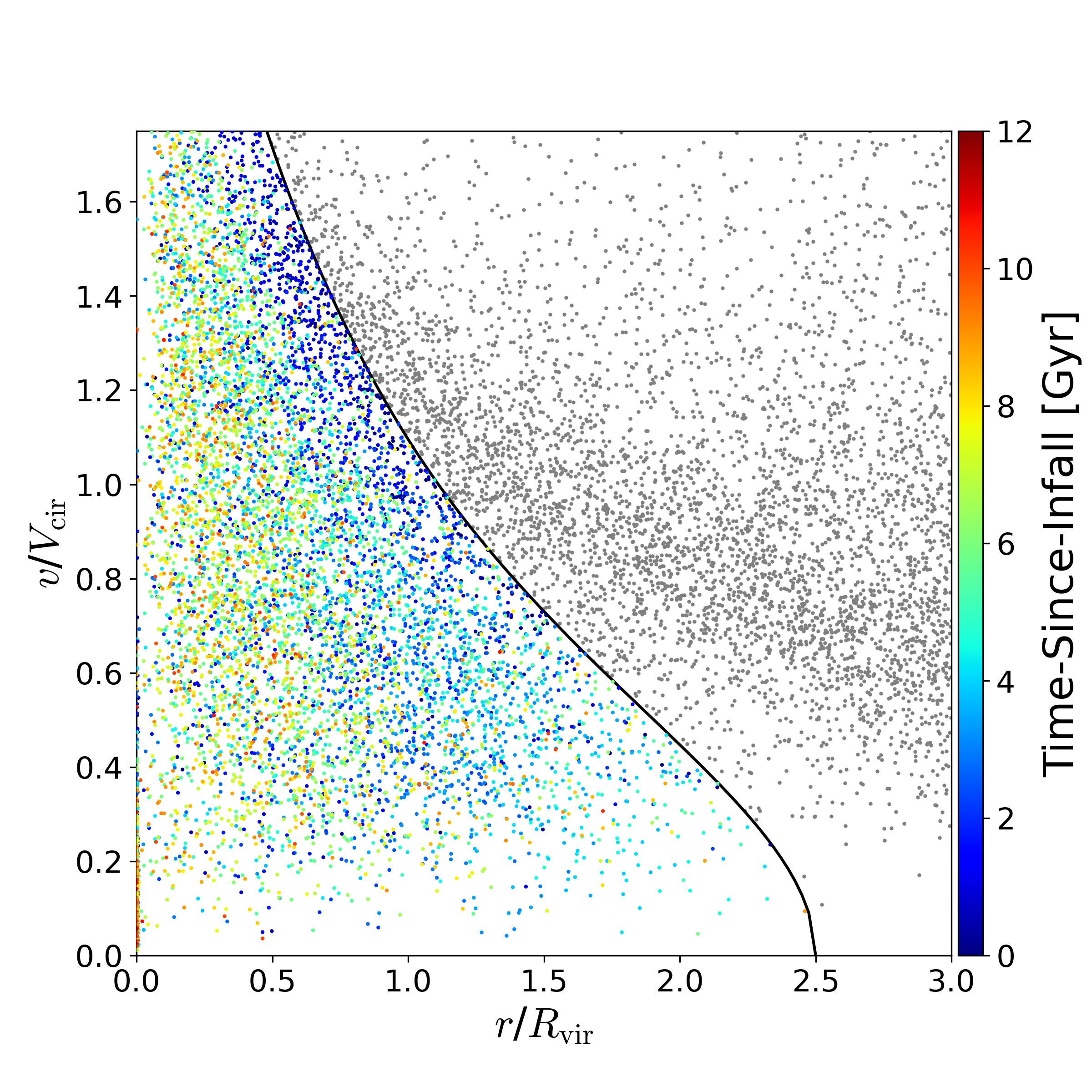}
\caption{
Phase space diagram of Horizon-AGN group/cluster member candidates in the last snapshot ($\rm z=0.018$).
The horizontal axis is the distance of the galaxy to the center of the host halo normalized by the virial radius of the halo.
The vertical axis is the relative velocity of the galaxy normalized by the circular velocity.
The black line indicates the relative velocity threshold of \cite{Han2018} for selecting satellites.
Grey points are galaxies not identified as members and other color scales show the time-since-infall of member galaxies (Section \ref{sec:Method-sample}).
}
\label{fig1_psd}
\end{figure}

\subsection{Sample Selection}
\label{sec:Method-sample}

The galaxies and halos in Horizon-AGN are identified using {\sc{HaloMaker}} which is built based on the AdaptaHOP algorithm \citep[][]{Aubert2004, Tweed2009}.
In total, 124,744 galaxies are identified in the last snapshot ($z=0.018$), and the stellar mass range is $10^{8.2-12.7} \msol$.
We use the galaxies with a stellar mass greater than $10^{9.5}$ \msol in the last snapshot to secure the minimum number of stellar particles (900) for reasonable kinematic stability \citep[see][for the issue of particle evaporation]{Klypin1999}.
This criterion leaves 53,226 galaxies as our final sample.

We classify the identified host dark matter halos with the final halo (virial) mass $\mhalo \geq 10^{13}\,\msol$ into ``group/cluster'' halos, as a conventional way \citep[][]{Dariush2010, Serra2013, Donnari2021b}.
We define the virial radius, $R_{\rm vir}$, as the radius within which the mean density is 200 times the critical density.
Note that the term ``halo'' refers to the main (group/cluster) dark matter halo that hosts multiple satellite galaxies in this study.
Using these mass criteria, 428 group/cluster halos were identified.
Subsequently, we define their member galaxies using the energy criterion introduced by \cite{Han2018}, i.e., by finding galaxies with negative orbital energy within 2.5 times the virial radius. After this cut, we obtain 8,228 group/cluster galaxies.
These member galaxies in the groups/clusters are our main focus, whilst the remaining galaxies are used as the control sample.
Among the rest, we exclude 3,797 galaxies that are within 3 $R_{\rm vir}$ but not bound to any group/cluster halo from our analysis because we want to highlight the difference between the field and group/cluster galaxies.
The remaining 41,201 galaxies are labeled as field galaxies.
Figure~\ref{fig1_psd} shows the phase space distribution of the member galaxies in the Horizon-AGN group/cluster halos at $z=0.018$.
The black line shows the relative velocity ($v$) threshold by the clustocentric distance ($r$) following \cite{Han2018}:
\begin{equation}
    \frac{v^2}{2} + \Phi(r) < \Phi(2.5 R_{\rm vir}),
\label{eq1_membership_criterion}
\end{equation}
where $\Phi$ is the gravitational potential of the group/cluster halo calculated from the radial density profile (i.e., NFW profile)\footnote{Note that the second condition in this equation was omitted in the original manuscript of \citet{Han2018}.}:
\begin{equation}
    \Phi(s) = \left\{
        \begin{array}{l l}
            {-V_{\rm cir}^2} \biggl[ h(c) g(c) \bigl( \frac{\ln(1+cs)}{s} - \ln(1+c) \bigr) \\ \qquad -\frac{1}{2}(1-s^2)(h(c)-1)+1 \biggr],  & s<1\\
            {-V_{\rm cir}^2} /s,  & s\geq1
        \end{array}
        \right.
\label{eq2_potential}
\end{equation}
where $s$ is the clustocentric distance of each galaxy in units of $R_{\rm vir}$ of the host halo ($r/R_{\rm vir}$), $V_{\rm cir}$ is the circular velocity at $R_{\rm vir}$ of the halo ($\sqrt{GM_{vir}R_{vir}}$), and $c$ is the concentration index of the halo assuming the NFW profile.
Other terms are defined as follows:
\begin{eqnarray}
    h(c) &=& \left[ 1- \frac{c^2 g(c)}{3(1+c)^2} \right]^{-1}
    \label{eq3_hc}\\
    g(c) &=& \left[ \ln(1+c)-\frac{c}{1+c} \right]^{-1}
    \label{eq4_gc}
\end{eqnarray}

Figure~\ref{fig1_psd} shows the galaxies within $3R_{\rm vir}$ of our sample group/cluster halos.
The member galaxies are color-coded by ``time-since-infall (TSI).''
Inspired by previous studies \citep{Wetzel2013, Oman2013, Rhee2017, Rhee2020}, we define time-since-infall (TSI) by measuring the time since the galaxy satisfies the aforementioned membership criterion, Equation~(\ref{eq1_membership_criterion}), in massive halos ($\mhalo > 10^{13}\,\msol$) for the first time.
While there are 787 snapshots in Horizon-AGN with intervals of $\sim\,17$ Myr, dark matter and gas data are saved only in 61 snapshots with time intervals of $\sim\,250$ Myr \citep[See also][]{Khim2020}.
As we use the dark matter snapshots to measure the TSI, it is therefore determined with a time resolution of $\sim\,250$ Myr.
The other galaxies are shown in grey.
Member galaxies appear roughly separated in phase space by TSI, which is consistent with the results of \cite{Rhee2017}.
We can see that this criterion saves the galaxies that are temporarily located out of $R_{\rm vir}$, although they are bound \citep[``back-splashed galaxies'';][]{Gill2005}.

\begin{figure*}
\centering
\includegraphics[width=0.9\textwidth]{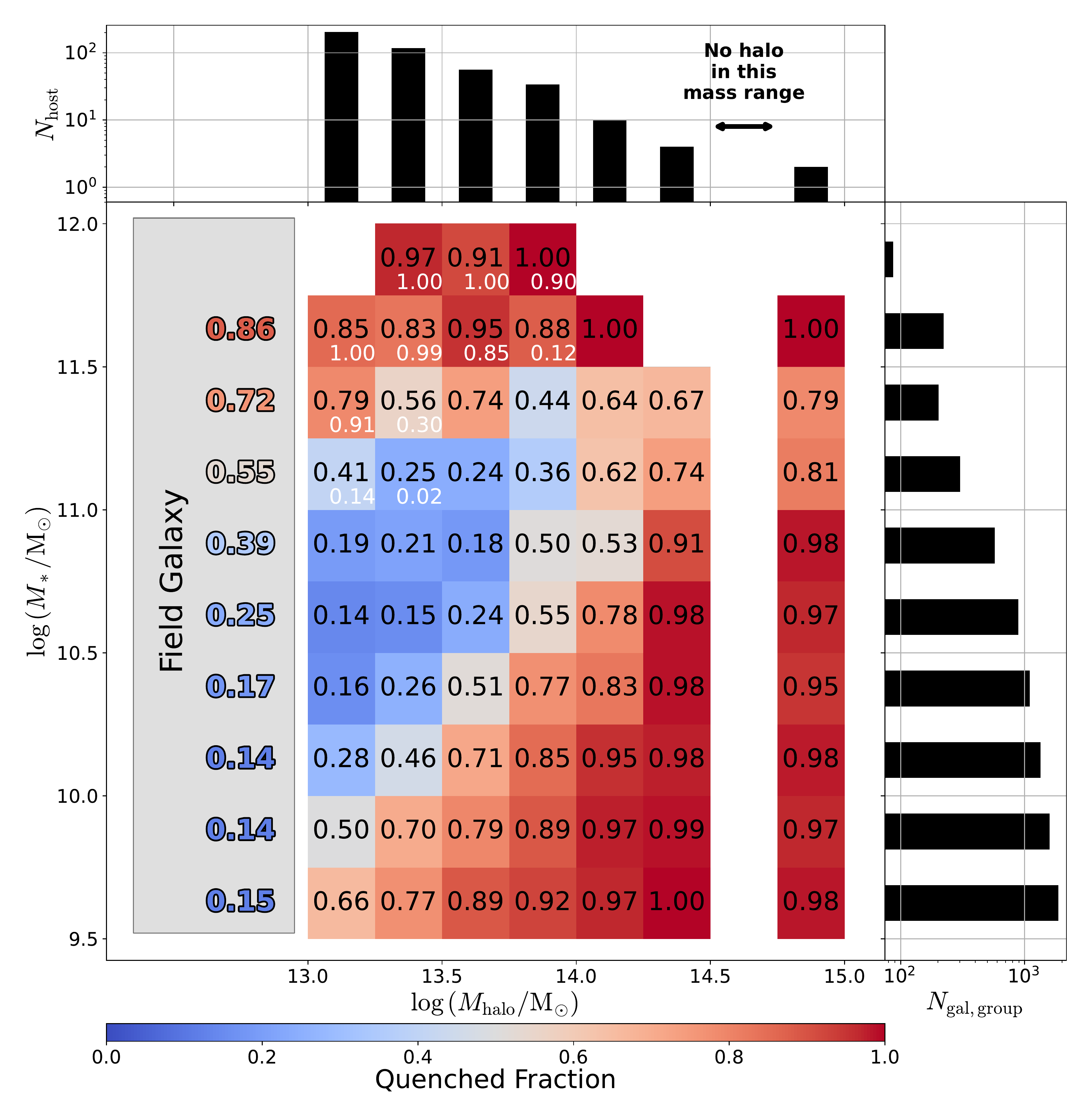}
\caption{
Quenched fraction distribution of the Horizon-AGN galaxies.
The horizontal axis is the host halo mass, and the vertical axis is the stellar mass of member galaxies.
The black numbers are quenched fractions color-coded following the key in the bottom and the white numbers are central galaxy fractions.
The mean values of the quenched fractions of field galaxies are shown in the grey inset.
The top and right panels show the number distributions of the host halos and their member galaxies, respectively.
Note that there are only two halos (clusters) of mass $\mhalo \geq 10^{14.5} \msol$ because of the limited size of the volume of Horizon-AGN (140\,Mpc on a side).}
\label{fig2_qf_tile}
\end{figure*}

%

\section[]{Results}
\label{sec:results}

\subsection{Quenched Fraction}
\label{sec:results-qf}

The star-formation quenched fraction is observationally well constrained at least at low redshifts.
For example, it is known to strongly depend on the stellar mass and environment of galaxies \citep[e.g.,][]{Baldry2006, Peng2010, Wetzel2013}.
Massive galaxies seem to be more quenched regardless of their environment, whereas low-mass galaxies are more quenched in higher density regions \citep[][]{Baldry2006}.
Such a consensus on the quenched fractions in observations can be used to test the reliability of numerical simulations.
In this section, we measure the quenched fractions of our sample of Horizon-AGN galaxies and attempt to understand the impact of stellar and host halo masses on the quenched fraction.

Various methods have been used to measure quenched fractions:
double Gaussian fitting on color-magnitude distributions \citep{Baldry2004, Balogh2004}, color-color (e.g., UVJ) distributions \citep{Labbe2005, Wuyts2007, Williams2009}, specific SFR (sSFR) \citep{Kauffmann2004, Santini2009}, and distance to the star-forming main sequence \citep{Tacchella2016, Fang2018, Pillepich2019}.
\cite{Donnari2021a} have reported that the quenched fraction as a function of stellar mass significantly varies with different measurement choices.
Bearing such uncertainties in mind, we use sSFR to define quenched galaxies, that is, when $\rm{sSFR} < \frac{1}{6 t_{\rm H}(z)}$, where $t_{\rm H}(z)$ is the age of the Universe at the epoch of the target redshift ($z$), corresponding to $\sim 10^{-11} \rm yr^{-1}$ at the present epoch \citep[][]{Tacchella2019}.

Figure~\ref{fig2_qf_tile} shows the quenched fraction distribution for different stellar and host halo masses.
The galaxies are divided into twelve stellar mass bins each with a 0.25 dex size, and group/cluster galaxies are additionally divided into eight halo mass bins of size 0.25 dex as well.
The number distributions of the group/cluster halos and member galaxies are shown in the top and right panels respectively, with the quenched fractions of field galaxies shown in the left inset.
Field galaxies, which are hardly affected by environmental effects, show a gradually increasing quenched fraction with increasing stellar mass.
This is often known as the ``mass quenching'' i.e., more massive galaxies exhibit a higher quenched fraction.
A relatively early and short SFH, perhaps associated with feedback effects, is often considered an important process behind mass quenching.
Furthermore, we notice three features of the group/cluster galaxies.
Firstly, the most massive members ($\mstellar > 10^{11.5}\,\msol$) are virtually all quenched regardless of the host halo mass.
This is similar to what has been found for field galaxies and consistent with observations \citep[][]{Peng2010, Wang2017, Wang2018}.
Hence, it seems that {\em the most massive galaxies are mass quenched whether they are in the field or in a cluster}.
We write the fraction of central galaxies in each bin in white letters. The most massive bins are dominated by central galaxies. But, whether including or excluding centrals makes little difference to the trend.
Secondly, the satellite galaxies in the most massive halos ($\mhalo > 10^{14.5}\,\msol$) are also predominantly quenched, regardless of the galaxy stellar mass.
This implies that environmental effects dominate the quenching process in massive clusters.
Lastly, there is a diagonally-increasing trend in the quenched fraction (from top left to bottom right in the galaxy mass range of $\mstellar \rm < 10^{11}\,\msol$).
The quenched fraction is higher in a more massive halo and lower mass galaxy.
This combined trend is largely consistent with various observations \citep[e.g.,][]{Baldry2006,Peng2010} as well as with the theoretical expectation based on the ram pressure stripping mechanism \citep{Jung2018}.

The high quenched fraction in low-mass satellite galaxies may call for attention.
The non-monotonic trend of the quenched fraction with respect to the stellar mass is a natural feature of the diagonal trend combined with the mass quenching.
However, the observational view of this issue is still unclear.
For example, \citet[see their Figure 6]{Peng2010} reported that intermediate-mass galaxies in dense environments exhibit lower values of red galaxy fractions compared with those of the surrounding mass bins, which appears to be consistent with our findings in the Horizon-AGN galaxies.
However, \cite{Baldry2006} did not find such a non-monotonic mass trend, despite being based on similar data.
If the Horizon-AGN galaxies are deemed unrealistic owing to the high values of quenched fractions in the low-mass regime, this could be related to the so-called satellite overquenching problem.
\cite{Kimm2009} reported that low-mass satellite galaxies exhibit substantially higher quenched fractions than their observational counterparts in several semi-analytic models.
Moreover, some hydrodynamical simulations report a similar issue, perhaps due to resolution effects.
For example, \cite{Dickey2021} investigated resolution effects using the SIMBA and EAGLE simulations, and found that the choice of resolution causes significant differences in the star-formation properties of low-mass galaxies.
Interestingly, \cite{Donnari2021a, Donnari2021b} identified a larger variation in rather more massive galaxies from resolution tests in the TNG simulation.

In summary, Horizon-AGN reproduces the two most robust observational facts: mass quenching at the high-mass end regardless of the environment and severe environmental quenching in massive clusters regardless of galaxy stellar mass.
The diagonally-increasing trend in the quenched fraction from massive galaxies in low-mass halos to low-mass galaxies in massive halos can be attributed to the ram pressure stripping phenomenon; however, determining whether it is consistent with observations requires further investigation.

\begin{figure}
\centering
\includegraphics[width=0.45\textwidth]{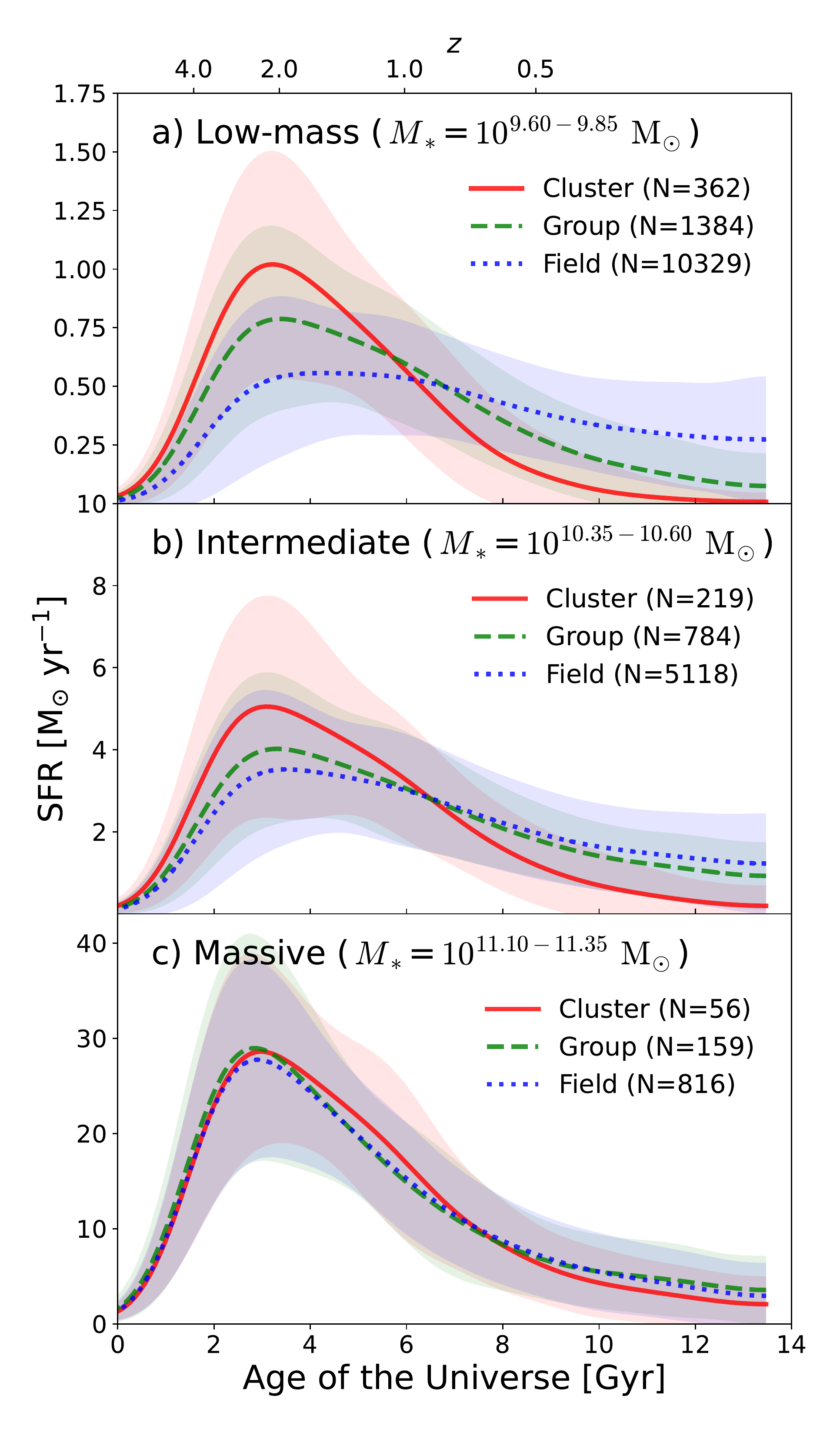}
\caption{
Mean SFHs of the Horizon-AGN galaxies of different masses.
Other mass bins not shown here are presented in Appendix~\ref{sec appendix_SFHall}.
Different colors and line styles represent the different classifications of environments, and the number of galaxies for each subsample is indicated in between brackets.
The group galaxies (green) mean the galaxies in the host halos of mass $10^{13-14} \msol$ and the cluster galaxies (red) are in the halos more massive than $10^{14} \msol$.
The shaded area shows the 1$\sigma$ standard deviation.
All the lines are smoothed using a Savitzky-Golay filter and a Gaussian kernel.
}
\label{fig3_sfh_halomass}
\end{figure}

\subsection[]{Star formation history}
\label{sec:results-sfh}

We have demonstrated that Horizon-AGN reproduces the observed quenched fractions of galaxies well at a given snapshot.
Now we focus on the entire SFH of galaxies and their relationship with stellar mass and host halo mass.
There are two conventional methods for building the SFH of a simulated galaxy.
One is to track the SFH of the main progenitors in the merger tree \citep[][]{Pandya2017, Iyer2017, Zheng2021}.
An advantage of this method is that it helps directly compare with observations at various redshifts because progenitors are observationally identifiable at various redshifts.
Another method is to build the SFH based on the age distribution of stellar particles which belong to the galaxy at the last snapshot \citep[][]{Sparre2015, Diemer2017, Joshi2021}.
This allows for the fact that in-situ formed stars may leave the host galaxy while ex-situ formed stars may join the galaxy later.
The star particle age distribution approach therefore encodes more complete information about the SFH and the current composition of the galaxy, therefore we use this approach.

Figure~\ref{fig3_sfh_halomass} shows the SFHs of the Horizon-AGN galaxies in different mass ranges.
The top panel shows the case of low-mass galaxies, the shapes of their SFHs are substantially different from each other due to their environments.
Cluster galaxies (the red line) are more star-forming at the beginning but quickly quenched.
On the other hand, field galaxies (the blue line) exhibit more extended star formation over time.
The bottom panel shows the massive galaxy case where there is little difference between group/cluster and field galaxies.
These results are consistent with previous observational studies \citep[][]{Thomas2005, Ellison2018}, and moreover, the differences in SFRs near the last snapshot also agree with the results of the quenched fraction (Figure~\ref{fig2_qf_tile}).
The mass ranges which are missing in Figure~\ref{fig3_sfh_halomass} show a continuous trend among the presented mass bins (see Appendix~\ref{sec appendix_SFHall}).
Thus, the environment significantly affects the SFH of galaxies in the case of low-mass galaxies.
When the SFR is measured at a certain epoch (e.g., the present epoch), it is not likely a transient feature but a result of the whole SFH containing the mass quenching prior to the infall into the host cluster and the environmental quenching after the infall.
When the mean properties are compared, cluster galaxies have been more passive than field galaxies for a substantial period (see the top panel of Figure~\ref{fig3_sfh_halomass}).
We call the point at which they cross the ``transition epoch'' and attempt to quantify it in the following sections.

\begin{figure}
\centering
\includegraphics[width=0.45\textwidth]{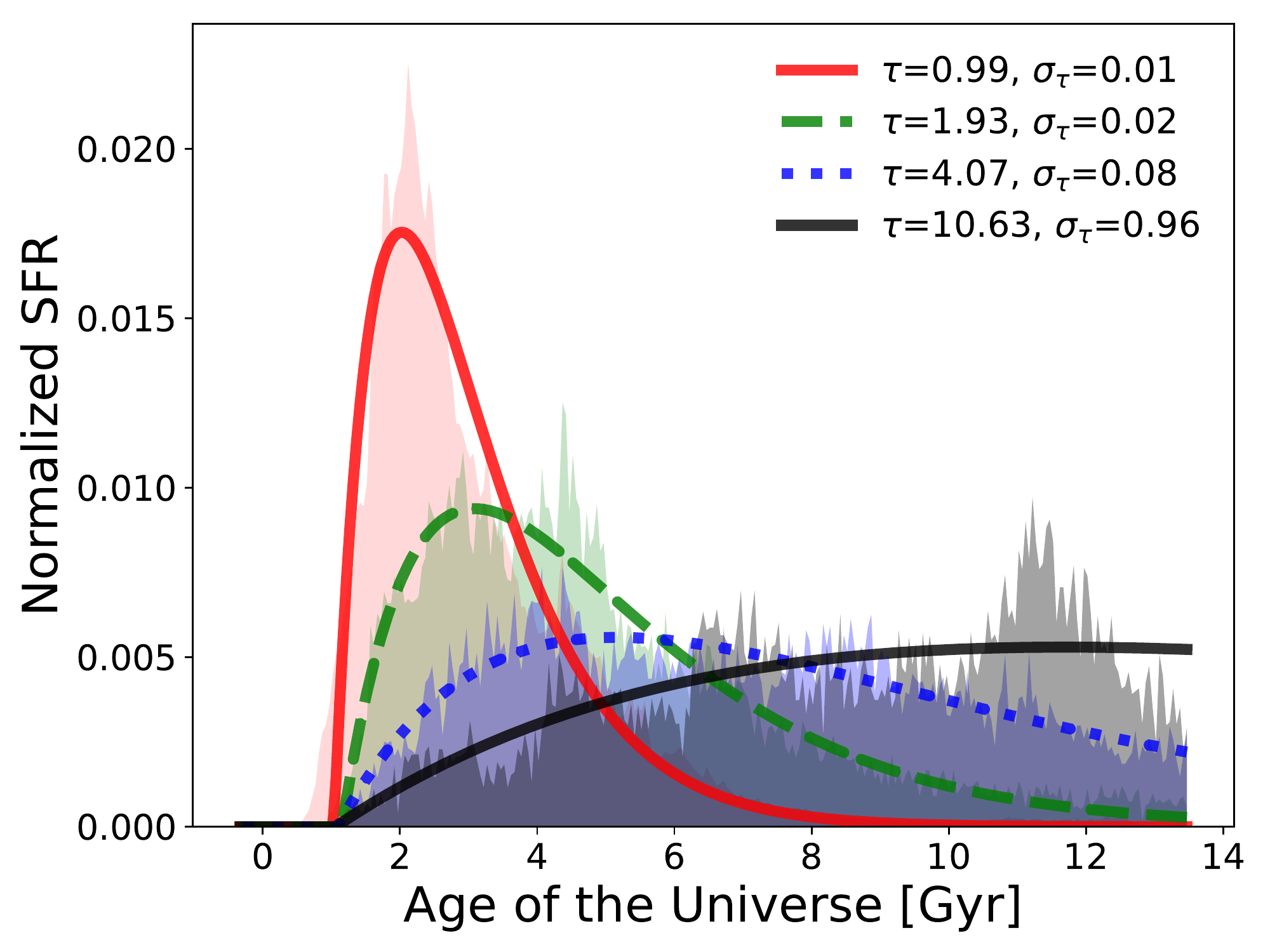}
\caption{
SFHs (the shaded area) of four sample Horizon-AGN galaxies and fits using Equation~(\ref{eq5_taumodel}) (the lines).
The different colors and line-styles represent the cases of $\tau \sim\ $1,\,2,\,4,\,and\,10.6 Gyr, respectively.
Each line and shaded area are normalized to the integrated SFR for clarity.
}
\label{fig4_SFH_example}
\end{figure}

\subsection{\elower-folding Timescale}
\label{sec:timescale}

The SFH of a galaxy may fluctuate with different timescales due to various physical processes.
However, environmental quenching, which is our focus, is rather a long-term process.
Thus, we aim to derive simple parameterized functions over time that fit the overall shape of the SFH.
We select the ``delayed tau-model'' \citep[see][ for reference]{Wetzel2013,Rhee2020} in this study because it is intuitively consistent with the expectation from simple closed-box systems \citep[e.g.,][]{Schmidt1959} yet versatile enough to allow a delay in the onset of star formation.
The following equation is used:
\begin{equation}
    \psi(t) = \left\{
        \begin{array}{l l}
            A(t-T_0) \exp{\bigl(-\frac{t-T_0}{\tau}\bigr)} & \quad t>T_0 \\
            0 & \quad \text{otherwise,} \\
        \end{array}
        \right.
\label{eq5_taumodel}
\end{equation}
where $\psi$ is the SFR, $t$ is the age of the Universe, $A$ is the amplitude, and $T_0$ is the delay in SFH.
We consider the e-folding timescale $\tau$ a proxy of the quenching timescale.
Figure~\ref{fig4_SFH_example} shows the schematic trends for various values of $\tau$ compared to a sub-sample of simulated galaxies.
The e-folding timescale, $\tau$, governs the overall shape of the SFH of a galaxy.
Moreover, as a larger $\tau$ represents more extended star formation, we can use this model to also fit galaxies with a rising SFR (see the black line in Figure~\ref{fig4_SFH_example}).

\begin{figure}
\centering
\includegraphics[width=0.45\textwidth]{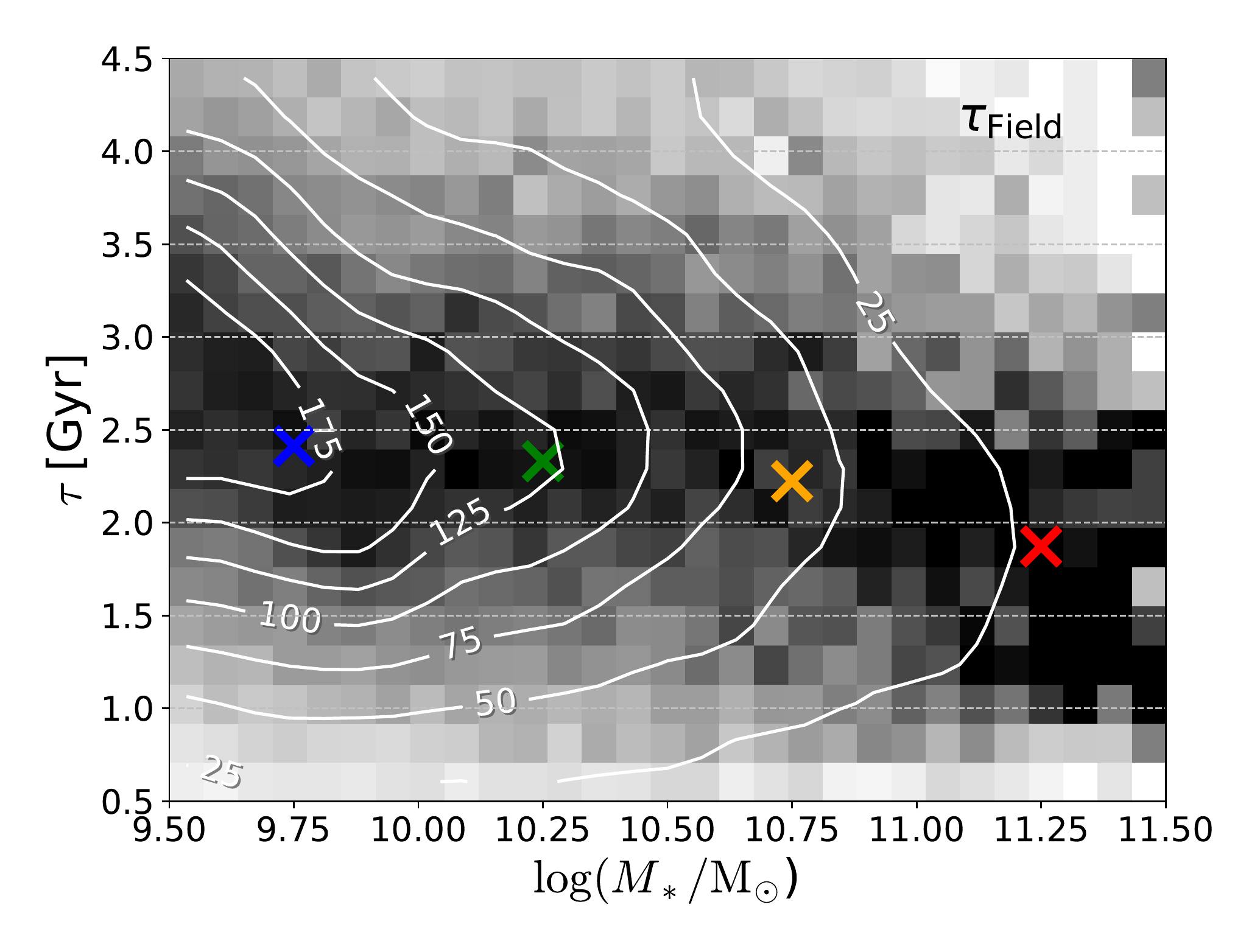}
\caption{
Distribution of the e-folding timescales ($\tau$) of field galaxies in the Horizon-AGN simulation.
The grey-scale histogram shows the column-normalized number density, and the white contours show the number of galaxies with labels marked.
The median values are presented for four mass bins.
}
\label{fig5_tau_field}
\end{figure}

\begin{figure}
\centering
\includegraphics[width=0.45\textwidth]{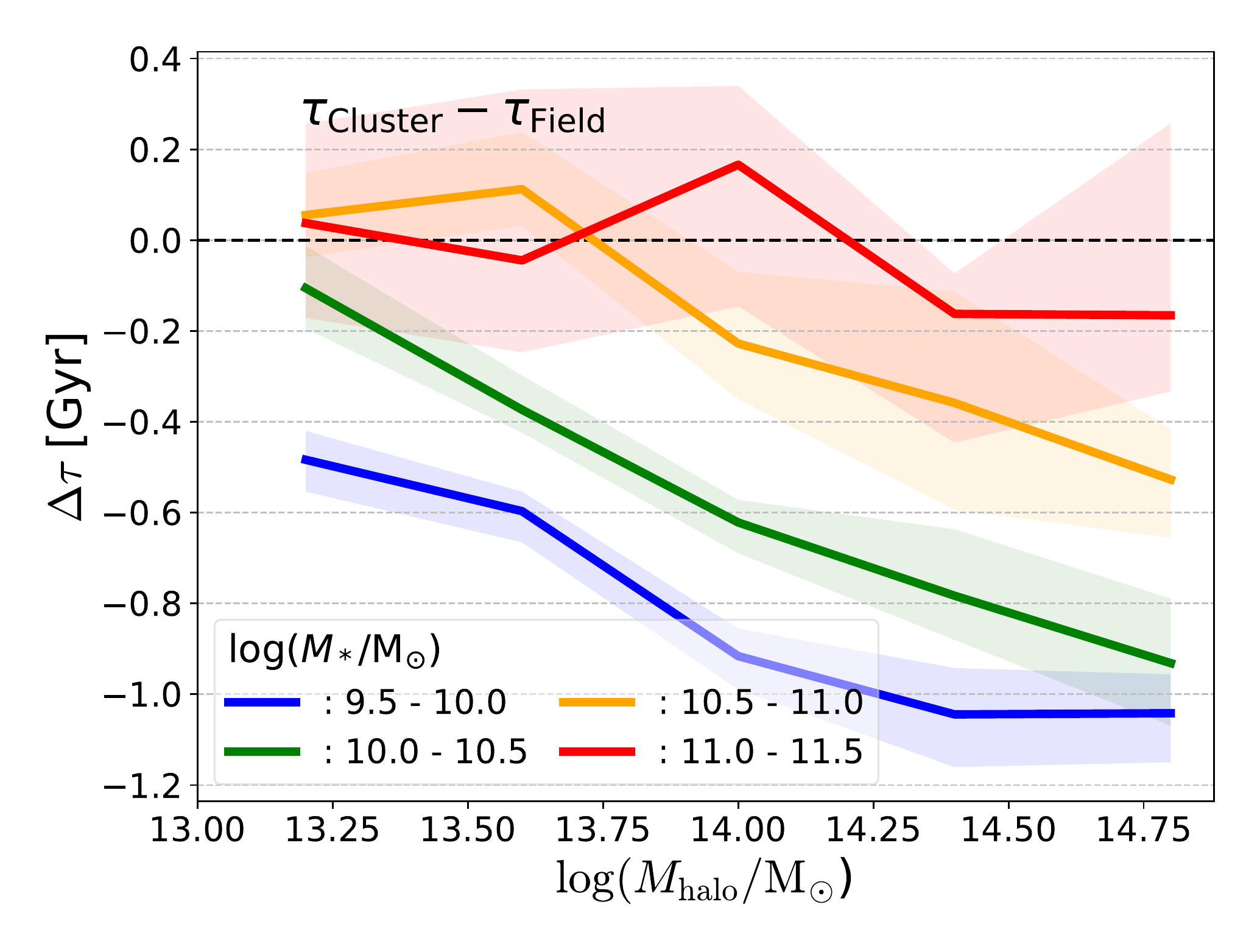}
\caption{
Residual values of $\tau$ (cluster galaxies - field galaxies) for different host mass bins, color-coded according to stellar mass bins in the same manner as illustrated in Figure~\ref{fig5_tau_field}.
The median value with its standard error of each bin is shown as a shaded line.
A small absolute value of $\Delta \tau$ indicates that the environmental effect to the SFH is negligible, while a negative value indicates the existence of star formation suppression by environmental effects.
}
\label{fig6_tau_residue}
\end{figure}

\begin{figure*}[!htb]
\centering
\includegraphics[width=0.9\textwidth]{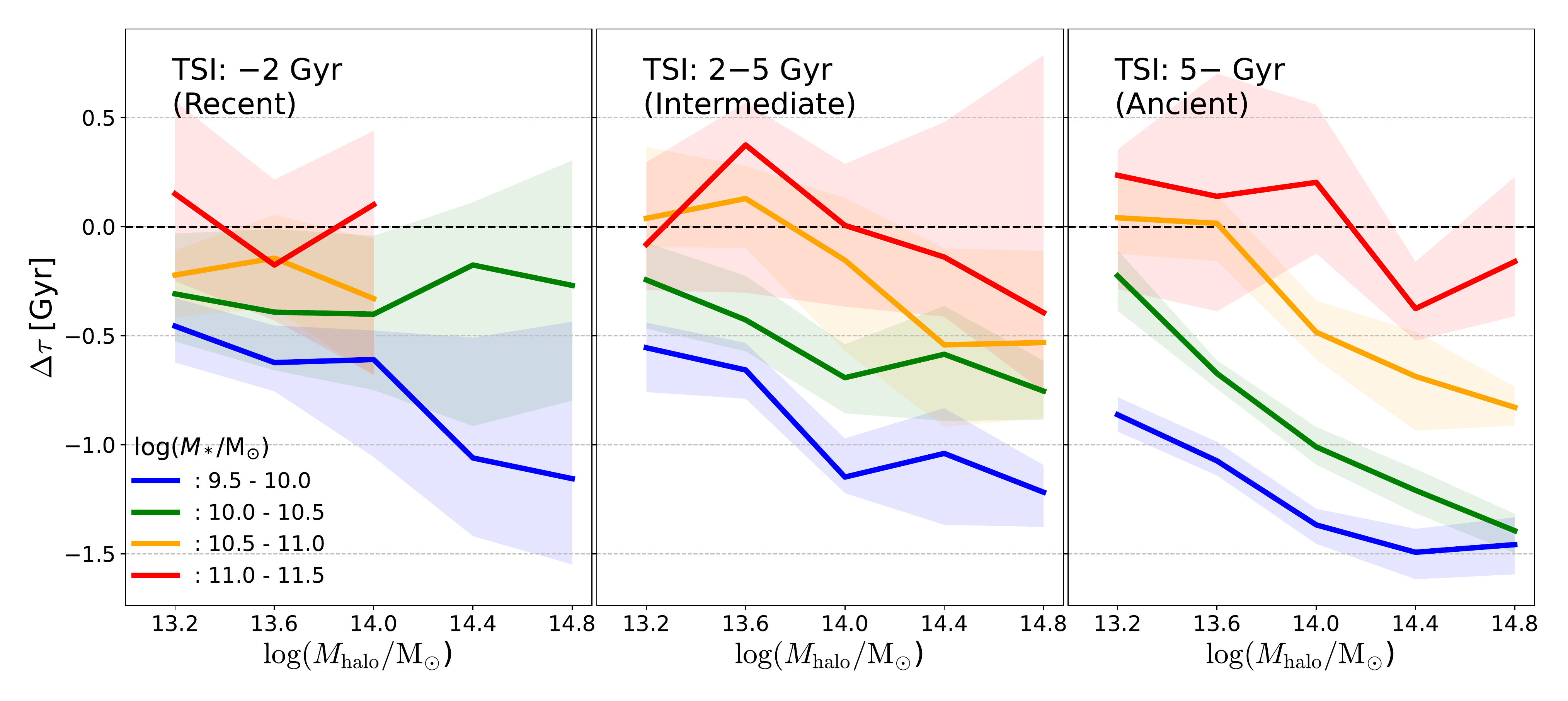}
\caption{
The residual values of $\tau$ as a function of the host mass, as in Figure~\ref{fig6_tau_residue} but divided into different TSI bins.
From the left to right, the TSIs of subsamples have been increased.
The color code is similar to the codes shown in Figures
~\ref{fig5_tau_field} and \ref{fig6_tau_residue} (the red and blue colors denote massive and less massive member galaxies, respectively).
}
\label{fig7_tau_residue_TSI}
\end{figure*}

Now we present how the $\tau$ of each galaxy depends on its stellar mass and the environment, as expected in Figure~\ref{fig3_sfh_halomass}.
Figure~\ref{fig5_tau_field} exhibits the values of the field galaxies in the Horizon-AGN simulation.
The white contours indicate number density, and the background grey-scaled histogram is the column-normalized number density of field galaxies in each bin.
The median values are color-coded for each stellar mass bin.
The e-folding timescales for the Horizon-AGN field galaxies range from 2.7 to 1.7, which agrees with the toy model results of \cite{Peng2010}.
There is a mild but systematic trend in which more massive galaxies exhibit smaller values of $\tau$, i.e., there is a rapid decline in SFH.
Consequently, more massive galaxies are more likely to be quenched early.
This explains the quenched fractions of field galaxies (Figure~\ref{fig2_qf_tile}).
Note that more than half of massive ($\mstellar > 10^{11}\,\msol$) field galaxies are quenched at $z \sim 0$.
Now, we take the field galaxies as a control sample to be compared with group/cluster satellite galaxies.

Figure~\ref{fig6_tau_residue} shows the difference in $\tau$ between the field and cluster galaxies ($\Delta \tau$).
A negative value of $\Delta \tau$ indicates that cluster galaxies exhibit smaller values of $\tau$, such that their star formation has experienced a more rapid decline than field galaxies for a fixed stellar mass.
Most satellite galaxies exhibit negative values of $\Delta \tau$,
this means that their SFRs have been further decreased by a massive halo environment.
This is particularly true for low-mass satellites ($\mstellar < 10^{10.5}\,\msol$) in more massive groups/clusters, where
additional environmental quenching appears to be more significant.
Conversely, the most massive member galaxies (red line) exhibit little difference ($\Delta \tau \sim 0$) from their field counterparts.
It seems that the most massive galaxies are less impacted by their environment.
These results are all consistent with the quenched fractions in Figure~\ref{fig2_qf_tile}.

It is clear that both stellar and cluster halo masses affect the quenching timescale.
However, not all cluster satellites exhibit an environmental effect if they have lived in a massive halo for only a short period of time.
Consequently, we want to check the significance of the time spent by a satellite galaxy in its host halo (group/cluster).
Herein, we adopt the time-since-infall (TSI) introduced in Section~\ref{sec:Method-sample}.
We separate galaxies into several TSI bins and recalculate $\Delta \tau$.
Figure~\ref{fig7_tau_residue_TSI} shows the satellite galaxies for three different ranges of TSI: recent, intermediate, and ancient infallers arriving into their host halos between 0 and 2\,Gyr ago, 2 and 5 Gyr ago, and 5 Gyr or more ago, respectively.
Most importantly, there is a clear trend with TSI.
The values of $\Delta \tau$ decrease from recent to ancient infallers, indicating the high significance of environmental effects on ancient infallers as expected.
However, it is somewhat unexpected that the environmental effect (significantly negative values of $\Delta \tau$) is clearly visible in low-mass galaxies despite the small values of the TSI (blue line in the left panel).
This phenomenon can be explained by the combination of two effects.
Lower-mass satellites are more easily affected by their host environment in the first place.
In addition, when they arrive in massive halos (e.g., $\mhalo > 10^{14}\,\msol$), they are likely to have experienced preprocessing quenching in smaller halo environments before the cluster infall \citep[][]{DeLucia2012, Han2018, Jung2018}.
Although it may not seem statistically significant, it is interesting to note that the most massive ancient infallers (red line in the right panel) exhibit negative values of $\Delta \tau$, hinting for additional environmental quenching.
In our simple quenched fraction analysis (Section~\ref{sec:results-qf}), and when we ignore the TSI difference (Figure~\ref{fig6_tau_residue}), these massive galaxies seemed to remain unaffected by environmental quenching.
However, when we delve into the TSI effect, we see a hint of additional environmental effects even in such massive galaxies.

In summary, we have derived the e-folding timescales ($\tau$) for the SFH of galaxies in the Horizon-AGN simulation.
The e-folding timescales exhibit a clear dependence on the galaxy mass and host halo mass.
The most massive galaxies seem to have an early and quick SFH almost regardless of the environment.
Except for the most massive galaxies (e.g., $\mstellar > 10^{11}\,\msol$), all the satellite galaxies exhibit environmental effects.
The magnitude of environmental effects seems larger in lower mass galaxies and in more massive halos, as expected from the ram pressure stripping process.
Moreover, we find that galaxies that have spent more time in their host halo have a shorter e-folding timescale, indicating a more significant environmental effect.
These findings on the simulated galaxies are largely in agreement with previous observational studies and consistent with basic physical expectations.

%

\section[]{Discussion}
\label{sec:Discussion}

\subsection{Transition epoch}
\label{sec:Discussion-transition}

\begin{figure*}
\centering
\includegraphics[width=0.9\textwidth]{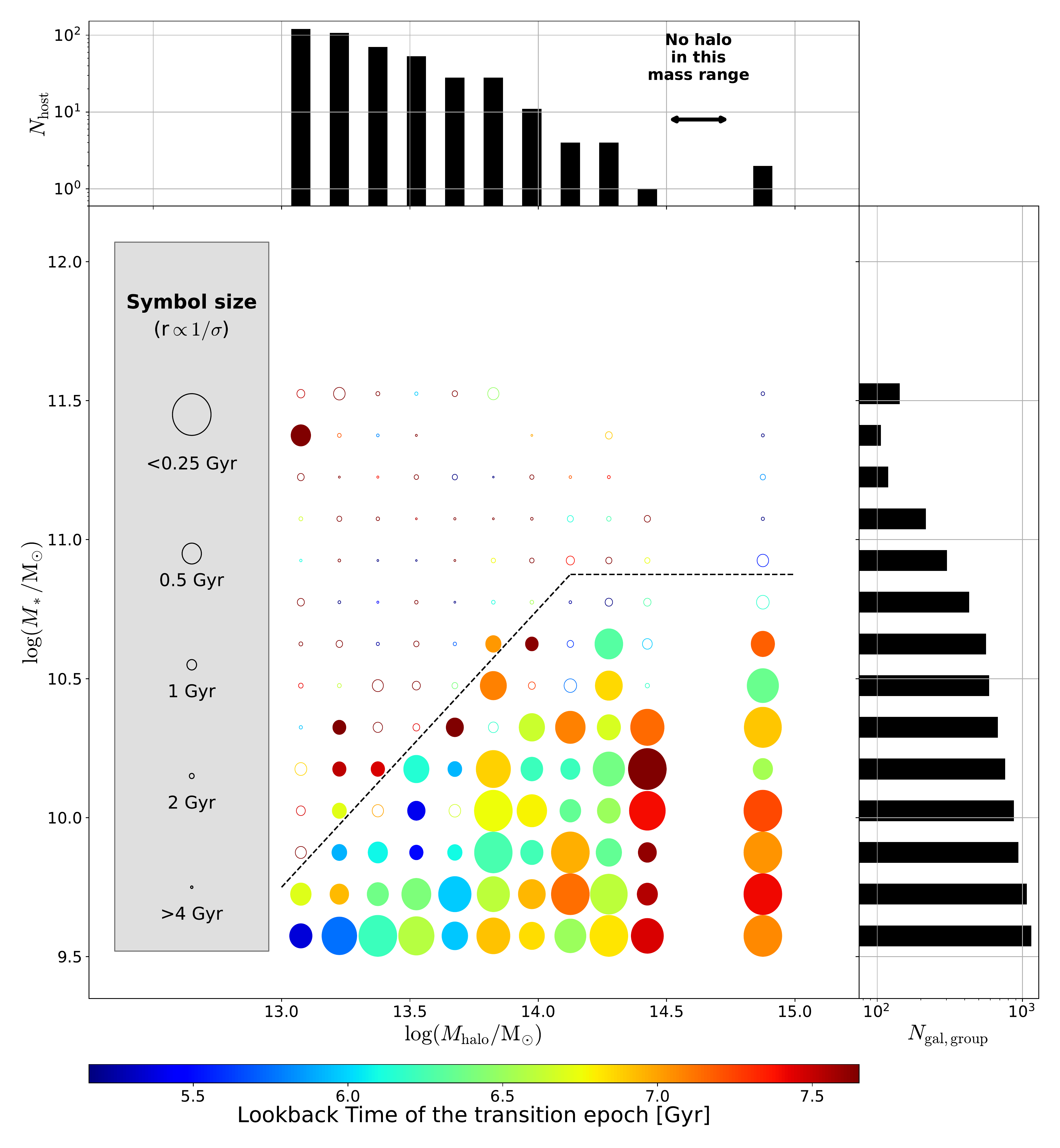}
\caption{
Transition epoch for the Horizon-AGN satellite galaxies in a bubble chart.
The horizontal axis is the host halo mass, and the vertical axis is the stellar mass of satellite galaxies.
The redder color means an earlier transition, and the bluer color means a later transition (color scales in the bottom).
We set the size of bubbles based on the distributional uncertainties of transition epochs.
Larger symbol size indicates that the transition clearly exists, whereas a small size indicates that there is no transition.
The size indicators are shown in the left inset, and we color-coded only bins that exhibited errors smaller than 0.75 Gyr.
The black dashed line is a guideline that distinguishes the reliability of the existence of the transition epoch.
The top and right panels show numbers of halos and galaxies shown in Figure~\ref{fig2_qf_tile}.
}
\label{fig8_transtion_bubble}
\end{figure*}

In the previous section, we established that today's ``red and dead'' cluster galaxies exhibited higher SFRs at an earlier epoch (Figure~\ref{fig3_sfh_halomass}~(a)) and smaller e-folding timescales of SFH decay than field galaxies.
It is natural that cluster galaxies switched from more active to less active at some point in history, compared with field galaxies \citep[][]{Gerke2007}.
We call this moment the ``transition epoch.''
Several studies have suggested similar concepts, that there was an epoch where the relation between SFR and density was reversed \citep[][]{Elbaz2007, Brodwin2013, Hwang2019}.
While the average SFR of galaxies decreases towards the center of a cluster at low redshifts \citep[][]{Balogh2000}, the trend becomes faint or even reversed at high redshifts.
We now attempt to determine whether the Horizon-AGN simulation presents a transition epoch as expected and, if so, whether it is universal or a function of the galaxy stellar mass or group/cluster host halo mass.
It is not trivial to uniquely determine the transition epoch in practice mainly because the actual SFHs of individual simulated galaxies are rather noisy, resulting in multiple crossings between the SFRs of the cluster and field galaxies.
We thus use the {\em cumulative} SFHs to mitigate complications.
A full description of the process is provided in Appendix~\ref{sec appendix_TEcalc}.
In addition, the reliability of an estimated transition epoch mainly depends on how significantly the SFHs of the cluster galaxies are different from those of their field counterparts.
Therefore, we perform bootstrapping to estimate the uncertainty in determining the transition epoch.

Figure~\ref{fig8_transtion_bubble} presents the results in a bubble chart color-coded by the measured transition epochs for each stellar mass and host halo mass bin, in a manner similar to that used for Figure~\ref{fig2_qf_tile}.
The size of the symbol represents the inverse of the dispersion in the transition epoch measured from bootstrapping; the larger the circle, the more reliable the measurement.
The transition epoch measurements exhibit small values of uncertainty ($< 0.75$\,Gyr, filled circles with color) in most low-mass galaxies and therefore are reasonably robust.
The first notable feature is that transition epochs are reliably measured only for relatively low-mass galaxies of $\mstellar < 10^{10.5} \msol$.
This is consistent with the results obtained from the e-folding timescales (Figure~\ref{fig6_tau_residue}).
More massive galaxies than that exhibit negligible environmental effects.
Next, we notice a diagonal trend from blue to red in the low stellar mass regime, i.e., as galaxy stellar mass decreases and as host halo mass increases, the transition epoch occurs earlier (larger values in lookback time).
This is again consistent with what was learned from the quenched fraction (Figure~\ref{fig2_qf_tile}).
We present the line for $\mratio =3.25$ in the figure to guide the eyes.
It appears that the transition epochs are clearly defined below this line.
This will be discussed further in the following section.

The transition epoch ranges from 5.2 to 7.6\,Gyr in lookback time depending on the stellar and cluster halo masses.
This corresponds to $z \approx 0.56$\,--\,1.02 according to the standard LCDM cosmology.
Considering the age of the Universe ($\approx 13.8$\,Gyr), this range of transition epoch (2.4\,Gyr or 17\% of the age of the Universe) may be viewed as narrow, in which case one may think that there is a ``universal'' transition epoch.
Whether we consider it universal or not, it may clearly be the case in which {\em the most apparent relation between the SFR and environment found in today's galaxies is valid only for half of the cosmic history}.
It would be interesting to robustly determine the transition epoch in terms of internal and external properties of galaxies in future observations.

\subsection{Halo-to-Stellar Mass Ratio Dependence}
\label{sec:Discussion-mass_ratio}

\begin{figure}
\centering
\includegraphics[width=0.45\textwidth]{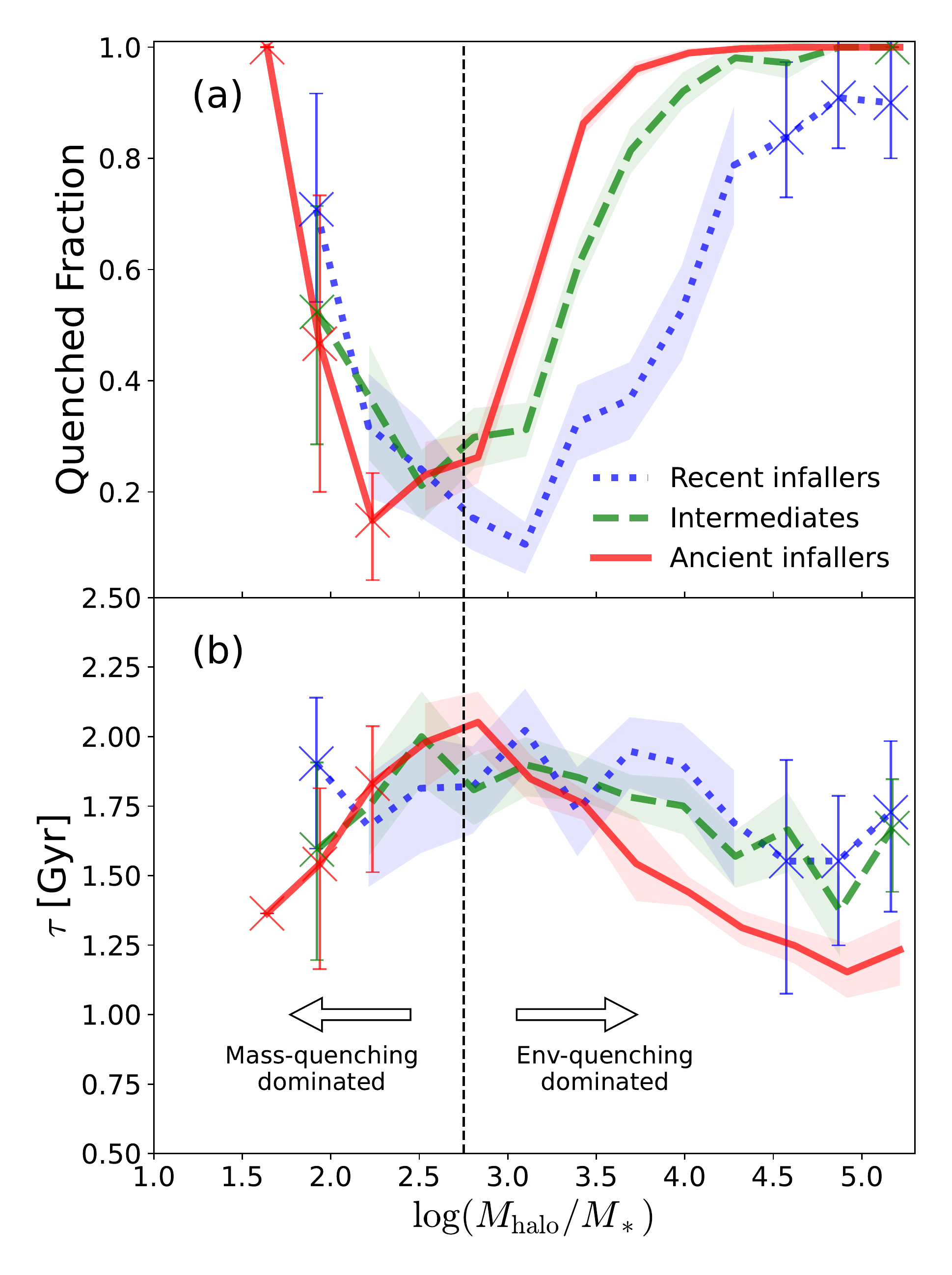}
\caption{
Upper panel~(a) shows the quenched fraction of the Horizon-AGN group/cluster galaxies as a function of the halo-to-stellar mass ratio.
The red line is the quenched fraction of the more ancient infaller galaxy, and the blue line is that of the more recent infaller galaxy.
The shaded areas show 95\% confidence intervals of the estimated quenched fractions.
The markers and the errorbars are same as the lines and the shades but the low number bins ($N_{\rm gal} \leq 50$).
The black dashed line at the $\mratio=2.75$ indicates the arbitrary demarcation where the trend is reversed.
The lower panel~(b) shows the results of $\tau$ as a function of the halo-to-stellar mass ratio for the same galaxies.
The color code and detailed format are the same as the upper panel.
}
\label{fig9_mratio}
\end{figure}

The main result of our analysis is that the quenched fractions and transition epochs are sensitive to both the stellar mass and host (cluster or group) halo mass: the ``diagonal trend.''
\cite{Baldry2006} expressed the quenched fraction using {\em a unified formula} that includes the environmental density and stellar mass (see their Equation~(10) and Figure~12).
Herein, we inspect the quenched fractions and e-folding timescales in terms of the halo-to-stellar mass ratio.

Figure~\ref{fig9_mratio}~(a) shows the quenched fractions of all satellite galaxies as a function of the cluster halo-to-stellar mass ratio.
We divide the satellite galaxies into three bins in TSI based on the same criteria used in Figure~\ref{fig7_tau_residue_TSI}.
For galaxies in halos with $\mratio \gtrsim 2.75$, a clear trend of the quenched fraction increasing with the halo-to-stellar mass ratio exists.
This seems to be the main condition for environmental effects to become strong.
However, galaxies with low halo-to-stellar mass ratios (i.e., $\mratio \lesssim 2.75$) exhibit an opposite trend.
This is a result of mass quenching.
These are mostly massive galaxies in small halos; hence, mass quenching is more effective than environmental quenching.
Moreover, ancient infallers in general show higher values of quenched fractions, which illustrates the significance of the TSI effect on the quenched states, as expected.

Similarly, we present the relation between $\tau$ and the mass ratio in Figure~\ref{fig9_mratio}~(b).
Panel~(b) is almost a mirror image of Panel~(a).
The environmental effect on the entire SFH of the galaxy is clearly visible at $\mratio \gtrsim 2.75$.
As the halo-to-stellar mass ratio increases, the values of $\tau$ go down to 1 Gyr.

\begin{figure*}
\centering
\includegraphics[width=0.9\textwidth]{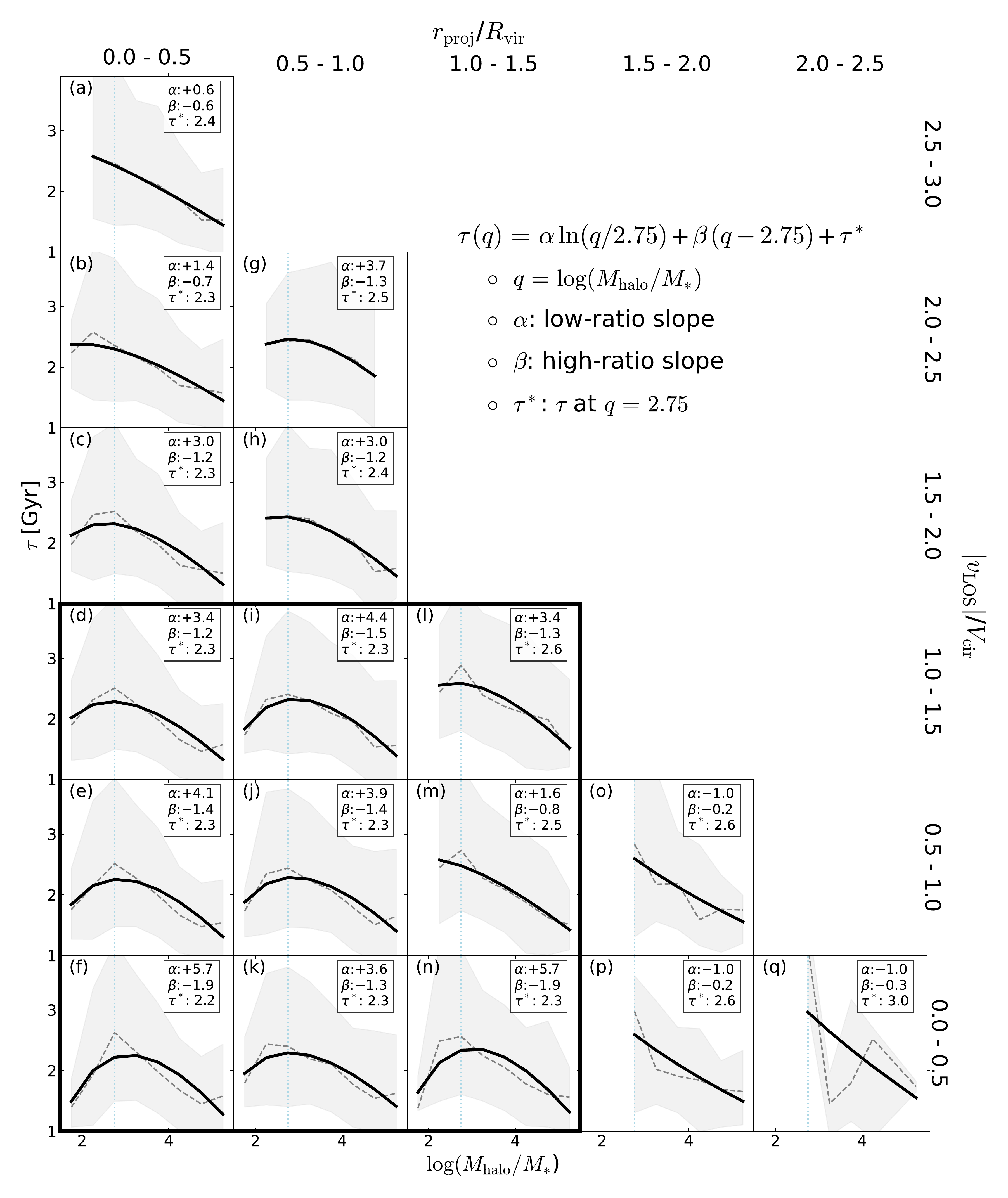}
\caption{
e-folding timescale as a function of the halo-to-stellar mass ratio.
Each panel shows the position on the two-dimensional phase space diagram.
The projected distance ($r_{\rm proj}$) and line-of-sight velocity ($v_{\rm LOS}$) of galaxies are measured from 300 random line-of-sights.
The grey dashed lines and shades show the median values and their \nth{16}-\nth{84} percentile uncertainties, and the black solid lines show the best fits to them.
The bins with less than 300 projected galaxies are not displayed.
The fitting parameters in Equation~(\ref{eq6_tau_empirical}) are given in each panel.
A relatively-more robust region of the cluster is roughly marked by the thick box that includes 12 panels.
}
\label{fig10_ratio_psd}
\end{figure*}

Based on these results, one may want to infer the e-folding timescale of SFH decay using the mass ratio and TSI.
However, TSI is difficult to determine for observed galaxies in reality \citep{Rhee2017}.
Hence, we derive $\tau$ for galaxies in various locations of phase space instead.
Figure~\ref{fig10_ratio_psd} presents the same information as in Figure~\ref{fig9_mratio}~(b) (i.e., the relation between the mass ratio and $\tau$) but sub-divided in different regions of projected position and line-of-sight velocity phase space.
To generate this diagram, we randomly generate 300 line-of-sight positions and velocities for each satellite galaxy following the exercise of \cite{Rhee2017} so that the observed data can be easily compared.
The phase space of this diagram goes beyond the virialized region marked by the curve in Figure~\ref{fig1_psd}.
We fit the data (grey dashed lines) using the parameterized formula (black curves):
\begin{equation}
    \tau(q) = \alpha\ln{(q/2.75)} + \beta(q - 2.75) + \tau_*
\label{eq6_tau_empirical}
\end{equation}
where $\tau$ is the e-folding timescale, $q$ is the logarithmic of the mass ratio (\mratio), $\alpha$ is the slope for the low-mass-ratio region, $\beta$ is the slope for the high-mass-ratio region, and $\tau_*$ is the characteristic e-folding timescale at $q=2.75$.
We first notice the $n$-shape in the fits of many panels, as shown in Figure~\ref{fig9_mratio}~(b).
The relatively more robust satellite members of the group/cluster halos are in the regions inside $r_{\rm proj}/R_{\rm vir} = 1.5$ and $|v_{\rm LOS}|/V_{\rm cir} = 1.5$, i.e., the nine panels marked by the thick black box.
Inside this region, the $n$-shape is apparent, and it is more apparent in the satellites close to the phase space center (i.e., closer to the halo center and with lower values of $|v_{\rm LOS}|/V_{\rm cir}$).
The slope of the fit for the high mass ratios ($\beta$) gradually increases from $-1.9$ in the central region of Panel~(f) to $-1.3$ in the outer region of Panel~(l).
This is all consistent with the results shown in Figure~\ref{fig9_mratio}~(b).
In the regions well outside the halos ($R_{\rm vir}>1.5$, Panels~(o) through (q)), there are no or few galaxies with mass ratios lower than the pivot value and therefore the fits are monotonic.
It is natural because such massive galaxies are likely to be brightest cluster/group galaxies that are rather centrally located in the halo.
The galaxies well outside the halos ($R_{\rm vir}>1.5$, Panels~(r) through (u)) exhibit significantly larger values of $\beta$, indicating a smaller environmental effect on them.
The fast-moving satellites in Panels~(a), (b), and (c) are likely recent infallers in group-size halos, which have not been sufficiently affected yet by environmental effects; hence they exhibit larger values of $\beta$ than those of the robust members.
The values of $\tau_*$ are remarkably similar among the panels, hinting for a universal demarcation between mass quenching and environmental quenching in the group/cluster galaxies.

%

\section{Conclusion}
\label{sec:Conclusion}

Using the hydrodynamical cosmological simulation, Horizon-AGN, we have investigated how the SFHs of galaxies vary with their internal and external properties.
We fit the SFH of each galaxy with a parameterized form and measure the e-folding timescale, $\tau$, which quantifies the pace of star formation quenching.
Distinct quenching features have been found in the Horizon-AGN galaxies, and our main results are as follows:
\begin{itemize}[leftmargin=*]

\item 
Massive galaxies exhibit higher values of star formation quenched fractions and smaller e-folding timescales in their SFHs than low-mass galaxies, regardless of their environments.
This implies that the current low star formation rates in massive galaxies are caused by a vigorous star-forming phase and subsequent rapid quenching in the early universe.
As expected, mass quenching, which is a dominant process for massive galaxies, is clearly seen in the Horizon-AGN galaxies.

\item 
Massive cluster environments additionally suppress the star formation of their member galaxies.
The galaxies in the most massive cluster halos exhibit significantly higher quenched fractions and substantial drops of $\tau$ compared with their field counterparts.
This environmental quenching becomes stronger in more massive halos.

\item 
The overall star formation activities, including the quenched fraction and the drop of $\tau$, exhibit a diagonal trend in the stellar mass versus halo mass plane.
The additional quenching of satellite galaxies fades when the host halo is not sufficiently massive enough to suppress the star formation of its satellites, which is consistent with theoretical expectations related to ram pressure stripping.

\item 
Time-since-infall (TSI), i.e., the total time duration of a satellite galaxy residing in a group/cluster halo, has a significant impact.
The longer a satellite galaxy resides in a group/cluster halo, the more it is affected by environmental quenching.
Ancient infallers with larger TSI values exhibit smaller values of $\tau$ than recent infallers.
Even the most massive satellites, whose SFH appears to be predominantly determined by mass quenching, show a moderate drop of $\tau$ when they have spent a long time in massive halos.

\item 
In Horizon-AGN, the transition epoch, where cluster galaxies become less star-forming than field galaxies, occurs 5.2\,--\,7.6 Gyr ago in lookback time.
This is largely consistent with the previous results of the ``reversal epoch,'' in which SFR-density relations are reversed \citep[][]{Elbaz2007, Hwang2019}.
However, the transition epoch can only be found for low-mass satellites, because massive galaxies do not show much dependence on the environment.
It ranges from 5.2 to 7.6 Gyr, displaying a diagonal trend on the stellar mass versus halo mass plane (low mass satellites in massive clusters exhibit early transition epoch), which is consistent with the trend in the e-folding timescale as well.

\item 
The cluster halo-to-stellar mass ratio and TSI may be key parameters for inferring the SFHs of satellite galaxies.
Satellite galaxies with different quenching processes are well separated by the mass ratio.
With an increase in the mass ratio, the quenched fractions gradually approach unity, and $\tau$ decreases to 1 Gyr.
This trend is reversed at $\mratio \lesssim 2.75$ because, in that domain, mass quenching is more important than environmental quenching.
We detect a systematic variation in the relation between $\tau$ and the mass ratio of the satellite galaxies in the observable phase space.

\end{itemize}

Through this investigation, we have attempted to understand the origin of the star formation properties of galaxies with respect to stellar mass and environment.
Horizon-AGN galaxies reproduce the most fundamental features such as mass quenching and environmental quenching.
Mass quenching is probably a result of stellar and AGN feedback.
A more massive galaxy forms stars more efficiently and consumes cold gas in the early part of cosmic history.
The mass of the central black hole scales with the dynamical mass of the galaxy, and thus, the feedback energy from the black hole must play a gradually more important role with increasing stellar mass \citep[][]{Volonteri2016}.
The pattern of environmental quenching seems consistent with the expectation from the ram pressure stripping mechanism.
As a result of collaboration or competition between the two quenching mechanisms, galaxies do develop and exhibit different properties at different epochs to the extent that some of the most apparent relations are reversed.
We are encouraged by the level of detail of the models that reproduce some key properties of present-day galaxies and provide predictions for the high-redshift Universe.

\begin{acknowledgments}
\section*{Acknowledgments}
We thank the referee for constructive criticism that clarified a few important issues in the original manuscript.
S.K.Y. acknowledges the support from the Korean National Research Foundation (2020R1A2C3003769).
A.C. acknowledges the support from the Korean National Research Foundation (2022R1A2C100298211).
T.K. was supported by the National Research Foundation of Korea (2020R1C1C1007079).
R.A.J. was supported by the Yonsei University Research Fund (Yonsei Frontier Lab, Young Researcher Supporting Program) of 2021.
This study was partially supported by the Spin(e) grant ANR-13-BS05-0005 of the French Agence Nationale de la Recherche and  the Center for Galaxy Evolution Research 
(2022R1A6A1A03053472).
It also relied on the HPC resources of the Horizon Cluster hosted by the Institut d’Astrophysique de Paris.
We warmly thank S. Rouberol for running the cluster on which the simulation was post-processed.

\end{acknowledgments}


\bibliography{ref}{}

%

\appendix
\restartappendixnumbering

\section{Star formation histories of entire samples}
\label{sec appendix_SFHall}

\begin{figure}[h]
\centering
\includegraphics[width=0.5\textwidth]{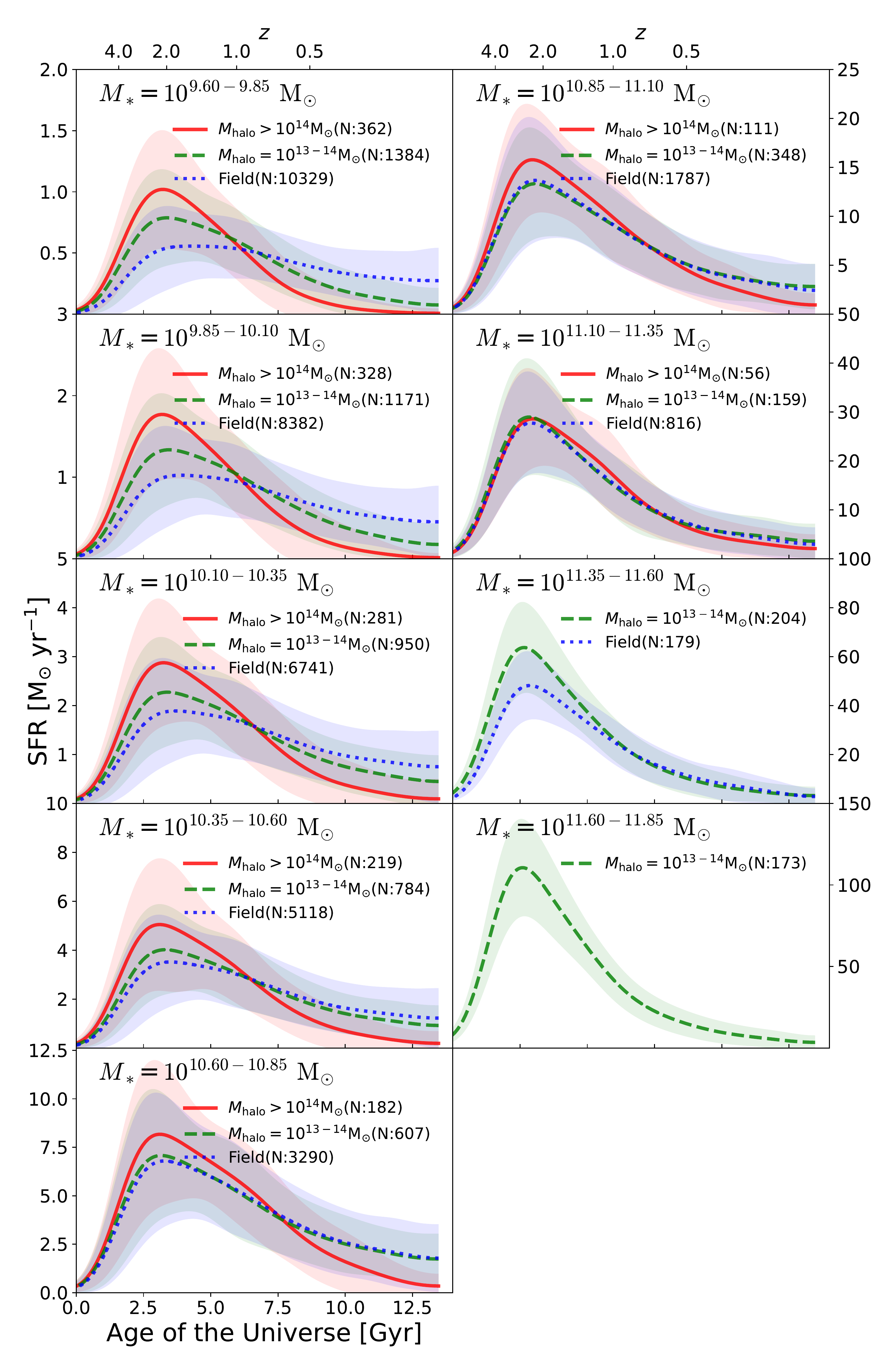}
\caption{
Mean SFHs of the Horizon-AGN galaxies of different masses in the same manner as Figure~\ref{fig3_sfh_halomass}.
}
\label{fig11_SFHall}
\end{figure}

We exhibited the mean SFH of Horizon-AGN galaxies in Figure~\ref{fig3_sfh_halomass} for representative stellar mass bins.
Here we present the same results of all mass bins in Figure~\ref{fig11_SFHall}.

\section{Transition epoch determination}
\label{sec appendix_TEcalc}

\begin{figure}[h]
\centering
\includegraphics[width=0.45\textwidth]{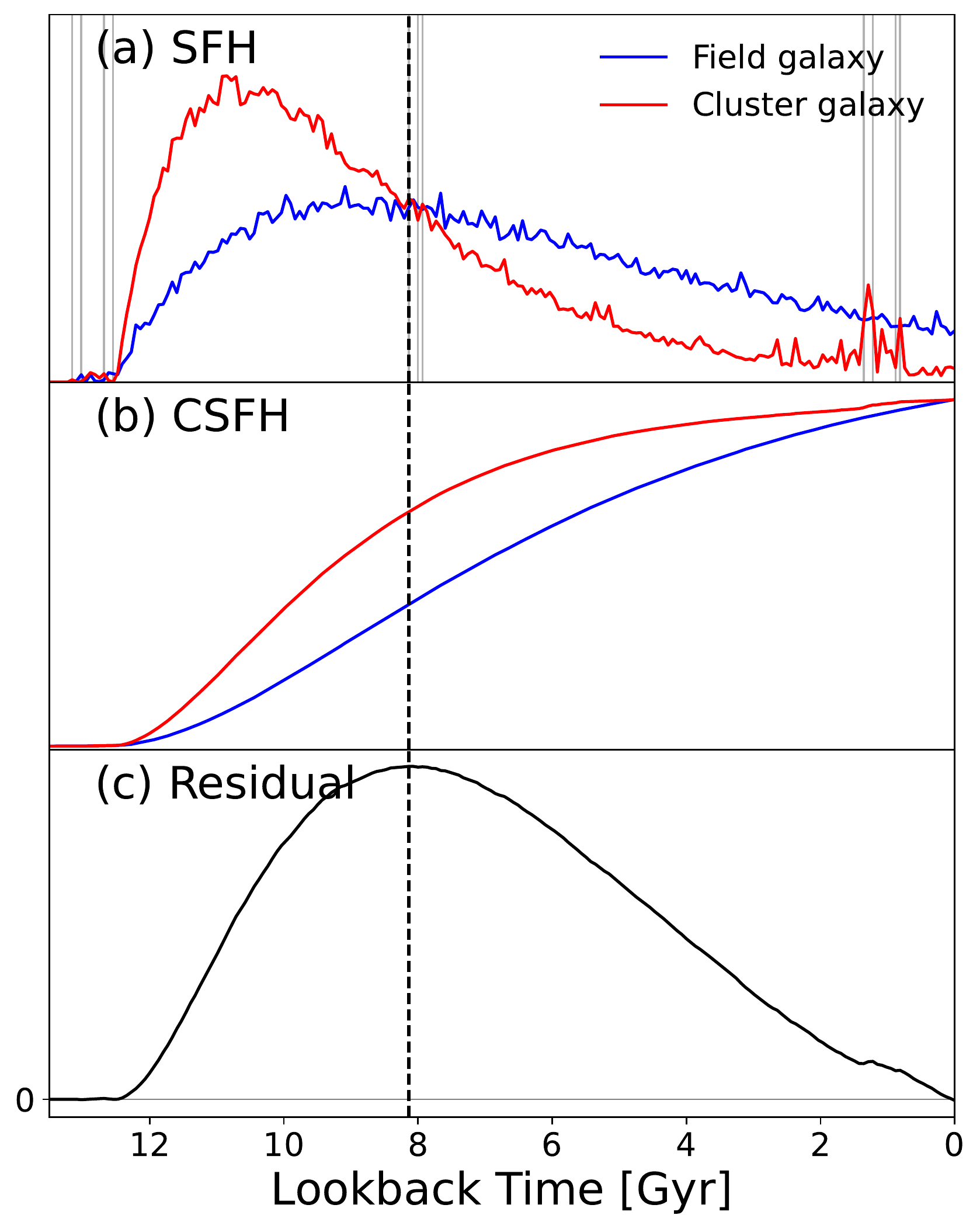}
\caption{
Procedure to determine the transition epoch.
We generate two mock SFHs with different $\tau$ considering their intrinsic noise.
The red and blue lines show the mock SFHs of cluster and field galaxies, respectively.
Panel~(a) shows the mean SFH of both samples.
Panel~(b) shows cumulative SFHs.
Panel~(c) presents the residuals of the two cumulative SFHs.
The grey vertical lines in Panel~(a) show all the crossing points that could be misidentified as the transition epoch.
The black dashed line shows the determined transition epoch, corresponding to the maximum point of the residual.
}
\label{fig12_TEcalc}
\end{figure}

We determine the {\em{transition epoch}} as follows.
First, we separate our simulated galaxies into stellar mass bins of size 0.25 dex and member galaxies into another 0.25 dex bins in host halo mass.
We derive the mean SFH of field and cluster galaxies in each bin.
A simple solution would then be to determine the intersection point of the SFHs of the field and cluster galaxies.
However, as star formation is often bursty and SFHs are very noisy, we decide to use cumulative SFHs.
We then calculate the residual cumulative SFH between cluster and field galaxies.
We regard the epoch when these residual values reach a maximum as the transition epoch.
The procedure is schematically illustrated for models in Figure~\ref{fig12_TEcalc}.
Furthermore, we can obtain the uncertainty of the estimated transition epoch through bootstrap sampling.
Based on 1000 bootstrapped pairs of SFHs, we estimate the mean and its error.

\end{document}